**Interface evolution in phase transformations ruled by nucleation and growth**

by

Massimo Tomellini

*Dipartimento di Scienze e Tecnologie Chimiche Università di Roma Tor Vergata*
*Via della Ricerca Scientifica 00133 Roma, Italy*

**Abstract**

An analytical model for the evolution of the boundary of the new phase in transformations ruled by nucleation and growth is presented. Both homogeneous and heterogeneous nucleation have been considered: The former includes transformations in 2D and 3D space and the latter nucleation and growth on flat solid substrate. The theory is formulated for the general case of spatially correlated nuclei, arbitrary nucleation rate and power growth law of nuclei. In the case of heterogeneous nucleation, spheroidal nuclei have been assumed and the dependence of the kinetics on contact angle investigated. The validity of the present approach is deemed through comparison with experimental data from literature which also comprise oxide growth by ALD (Atomic Layer Deposition) metal electrodeposition at solid substrate and alloy recrystallization.

## 1- Introduction

The synthesis and the design of solid systems with the desired interface morphology is a topic of great moment in Materials Science. This is due to the important role played by the interface whenever it is the locus, in a device, where chemical and physical processes take place. The term interfacial phenomena is intended here to include processes occurring at the solid-gas, solid-solid and solid-liquid interfaces. As a few examples, it is worth mentioning the adsorption in porous systems at the gas-solid interface, the 2D-interfaces in chemical sensors and the catalytic activity of supported metal catalysts where the chemical reaction occurs at the border between metal or oxide surfaces and gas phase [1, 2 ,3 , 4]. In gas sensors the interaction between gas molecules and materials mainly takes place on the surface, hence the number of surface atoms is crucial for determining the performance of the sensor [2].

In electrochemistry, the morphology of the liquid-solid interface is important for the electrodeposition of metals at electrodes via nucleation and growth. In fact, under potentiostatic conditions and in case of either interface or diffusion-controlled growths, the kinetics of metal deposition depends upon the evolution of the extension of the metal/liquid interface. It is the evolution of this interface that affects the shape of the cronoamperometric curve [5, 6, 7].

For the functionality of nano-composites, interfaces may play an important role for controlling the transport of ions, electron and phonons, either in matrix or through the filler. For instance, in dye-synthesized solar cells, photogenerated positive charges are normally considered to be carried away from the dyes by a separate phase of hole transporting material (HTM). It has been shown through experiments that the regeneration yield, ascribed to the hole diffusion, is linked to the extension of the interface between the $T_iO_2$ and HTM [8]. Besides, during the growth of tin-based alloy films the nucleation and growth of voids leads to the formation of interfaces whose extension is significant in controlling the properties of solder joints of microelectronic devices [9]. Another field that is currently attracting attention from researchers is that of thin oxide layer growth by thermal Atomic Layer Deposition (ALD). This highly scalable method is promising for using III-V semiconductors as replacement of Si in MOSFET transistors. The oxide growth rate in ALD is linked to the extension of the surface area of the oxide, which exhibits a maximum before impingement among islands becomes significant [10].

The present article is devoted to study the evolution of the interface extension of the new phase in transformations occurring by nucleation and growth. An analytical model has been developed which comprises both cases of homogeneous and heterogeneous nucleation. With respect to previously developed modeling, which are limited to random nucleation and constant rate of nucleus growth [7, 11,12], the present theory is formulated for the general case of spatially correlated nucleation, power growth law of nuclei and arbitrary nucleation rate.

The article is divided as follows: Section 2.1 is devoted to the definition of the probability functions that are needed in the stochastic approach developed in sect.2.2 for computing the kinetics of interface extension. For correlated nuclei these probabilities are given in terms of *n*-dots correlation functions. In section 2.3 the approach is applied to the model case of random nucleation with spherical and ellipsoidal-cap nuclei, for homogeneous and heterogeneous nucleation, respectively. The last two sections are devoted to numerical results and application of the model to describe experimental data.

## 2-Results and Discussion

### 2.1 Definition of the probabilities

We consider phase transformations ruled by nucleation and growth of spherical or circular nuclei in a homogeneous system. The growth law is given by the function $r(t, t')$, that is the radius of the nucleus[1], at time $t$, with $t' < t$ being the birth time of the nucleus. The nucleation rate of *actual* nuclei is indicated as $I_a(t)$ and the "phantom-included" nucleation rate (compatible with correlation constraints) as $I(t)$.

Let $d\xi$ be the probability that a generic point of the space, say c, is transformed between time $t$ and $t + dt$ by an actual nucleus (N) which start growing between $t'$ and $t' + dt'$ and located at relative distance $\boldsymbol{r}$ with $r = r(t', t)$ (Fig.1a). This probability can be expressed as [13],

$$d\xi = P_c(t|t', \boldsymbol{r})dP(t', \boldsymbol{r}), \qquad (1a)$$

---

[1] It is the radius for free growth, i.e. unimpeded by impingement with other nuclei.

where $dP(t', \boldsymbol{r})$ is the probability that an actual nucleus nucleates within $d\boldsymbol{r}$ around $\boldsymbol{r}$, with $r = r(t', t)$, and $P_c(t|t', \boldsymbol{r})$ is the conditional probability that the point c is untransformed at time $t$ provided that the nucleation of an actual nucleus occurred at $(t', \boldsymbol{r})$. It turns out that

$$dP(t', \boldsymbol{r}) = I_a(t')r^{D-1}(t', t)d_t r(t', t)d\Omega_D dt' . \qquad (1b)$$

where $D$ ($D$=2,3) is the space dimension, $\Omega_D$ the polar or solid angle and $d_t r(t', t) = \partial_t r(t', t)dt$. The probability $d\xi$ can also be rewritten by defining the conditional probability for the *nucleation event*, rather than for the point to be transformed, according to

$$d\xi = \big(1 - \xi(t)\big)dP_a(t', \boldsymbol{r}|t), \qquad (2a)$$

where $\big(1 - \xi(t)\big)$ is the probability the generic point c is untransformed at time $t$, and $dP_a(t', \boldsymbol{r}|t)$ the conditional probability that an *actual* nucleus nucleates at $(\boldsymbol{r}, t')$ given that the point c is untransformed up to $t$. In the case of diffusion type growth the last requirement guarantees that the nucleation event does not lead to the formation of a phantom which could imply (diffusion type growth) the point c to be transformed by an overgrowth phenomenon [14, 15]. The conditional probability $dP_a(t', \boldsymbol{r}|t)$ reads

$$dP_a(t', \boldsymbol{r}|t) = q(t', \boldsymbol{r}|t)I(t')r^{D-1}(t', t)d_t r(t', t)d\Omega_D dt' \qquad (2b)$$

where $q(t', \boldsymbol{r}|t)$ is the conditional probability that the nucleation point (at $\boldsymbol{r} = r(t', t)\hat{\boldsymbol{r}}$ within $r^{D-1}(t', t)d_t r(t', t)d\Omega_D$) is not transformed up to $t'$ (by any nucleus located at $r' > r(t', t)$) provided the point c is untransformed up to $t$ [15]. From now on we consider homogeneous and isotropic systems where, $P_c(t|t', \boldsymbol{r}) = P_c(t|t', r(t', t)) = P_c(t|t')$, $q(t', \boldsymbol{r}|t) = q(t'|t)$ and $\int d\Omega_D = \Omega_D = 2(D-1)\pi$. Under these circumstances eqns.1-2 implies

$$\frac{d\xi}{dt} = \big(1 - \xi(t)\big)\Omega_D \int_0^t q(t'|t)I(t')r^{D-1}(t', t)\partial_t r(t', t)dt'$$

$$= \Omega_D \int_0^t P_c(t|t') \, I_a(t') r^{D-1}(t',t) \partial_t r(t',t) dt' \quad . \qquad (2c)$$

Eqn. 2c gives the following relationship between the $P_c$ and $q$ probabilities,

$$P_c(t|t') = q(t'|t)\big(1 - \xi(t)\big) \frac{I(t')}{I_a(t')}. \qquad (3)$$

Also, the rate of transformation can be expressed in terms of actual quantities as,

$$\frac{d\xi}{dt} = \int_0^t I_a(t') \partial_t v_a(t',t) dt', \qquad (4)$$

where $v_a(t',t)$ is the mean actual volume of the nucleus[2]. From eqns.2c, 4 it follows that

$$\frac{\partial_t v_a(t',t)}{\partial_t v(t',t)} = P_c(t|t') \quad . \qquad (5)$$

Eqn.5 is the starting formula for computing the interface extension for homogeneous and heterogeneous nucleation. As reported in more detail in Appendix, the $P_c(t|t')$ probability can be expressed through a series in terms of $n$-dots (i.e. $n$-nuclei) correlation functions.

*2.2.Computation of the interface extension*

*2.2.1 Homogeneous Nucleation*

---

[2] To simplify the notation the subscript D will be omitted in the $v$ and $v_a$ volumes.

The computation of either the interface length ($D = 2$) or surface ($D = 3$) is done through eqn.5 by considering that $\partial_t v_a(t',t) = x_a(t',t)\partial_t r(t',t)$ where $x_a(t',t)$ stands for the length (or the area) of the interface of the single actual nucleus, and $\partial_t v(t',t) = x(t',t)\partial_t r(t',t)$ with $x(t',t) = \Omega_D r^{D-1}(t',t)$. It follows that $x_a(t',t) = P_c(t|t')x(t',t)$ and the total extension of the interface (per unit of $D$-dimensional volume) is eventually obtained by integration over the population of actual nuclei:

$$X_L(t) = \int_0^t I_a(t')P_c(t|t')x(t',t)dt'. \qquad (6)$$

*2.2.2 Heterogeneous Nucleation.*

In this section we consider nucleation on planar substrate of spheroidal nuclei, specifically prolate (oblate) ellipsoidal-cap. By denoting with $a$, $b$, and $c$ the semi-axes we get $a=b$ where $c$ is along the normal to the surface. The nucleus growth laws are $a(t',t)$, $c(t',t)$ where the aspect ratio ($c/a$) is constant during the growth.

For transformations on planar surface the computation of the transformed volume has been performed in ref.[16]. By denoting with $h$ the distance along the substrate normal, the volume per unit of substrate surface is given by

$$V(t) = \int_0^{h_{max}} S(t,h)dh, \qquad (7)$$

where $S(t,h)$ is the fraction of substrate surface covered by the new phase at height $h$. In ref.[16] it was shown that the integrand of eqn.7 describes a 2D-phase transformation with growth law $R(t',t,h)$ and nucleation law, $I_a(t,h)$. For instance, in the case of hemispherical nuclei it is $R(t',t,h) = [r^2(t',t) - h^2]^{1/2}$ with $r$ nucleus radius. The surface fraction is given by $S(t,h) = 1 - e^{-\sigma_{ex}(t,h)}$ where $\sigma_{ex}(t,h)$ depends on spatial distribution of nuclei and on both nucleation and growth rates. Application of eqn.5 to the $h$-dependent 2D-transformation leads to

$$\frac{\partial_t s_a(t', t; h)}{\partial_t s(t', t; h)} = P_c(t|t'; h) \ , \tag{8}$$

with $s_a(t', t; h)$ and $s(t', t; h) = \pi R^2(t', t, h)$ actual and extended surfaces of the section of the single nucleus at $h$, respectively.

In the following, we deal with the general case of ellipsoidal-cap nuclei with contact angle, $\varphi$ (prolate hemi-ellipsoid). A point on the surface of the ellipsoid is identified by $(x, y, z)$ coordinates (measured from the ellipsoid center) with $z$-axis along substrate normal (Fig.1b). In polar coordinates the equation of the spheroid is $\rho^2(1 - \epsilon\,(\sin\theta)^2) = a^2$, with $\epsilon = 1 - \frac{a^2}{c^2}$ a function of the eccentricity[3]. Moreover, the aspect ratio of the nucleus (and therefore the eccentricity and contact angle) are assumed to be constant during the growth (see below). Since $\rho = \frac{z}{\sin\theta}$ the equation of the spheroid is rewritten as $\rho^2 = a^2 + \epsilon z^2$ and the radius of the circular section, at $z$, becomes

$$R^2(t', t, z) = \rho^2(t', t, z) - z^2 = a^2(t', t) - (1 - \epsilon)z^2 \ . \tag{9}$$

The increment of the lateral surface of the actual nucleus in the layer at distance z reads

$$ds_{L,a}(t', t; z) = p_a(t', t; z)\frac{dz}{\sin\alpha} \ , \tag{10}$$

where $p_a$ is the length of the interface (perimeter of the nucleus) and $\tan\alpha = \left|\frac{dz}{dR}\right|$. Eqns.9 provides $\tan\alpha = \left(\frac{c}{a}\right)^2\frac{R}{z}$, that is $\sin\alpha = \frac{R}{\sqrt{(1-\epsilon)^2 z^2 + R^2}}$, and eqn.10 becomes

---

[3] For prolate and oblate ellipsoid $\epsilon = e^2$ and $e^2 = -\epsilon/(1 - \epsilon)$, respectively, where $e$ is the eccentricity. In the latter case $\epsilon < 0$.

$$ds_{L,a}(t',t;z) = \frac{p_a}{R}\sqrt{(1-\epsilon)^2 z^2 + R^2}\,dz \qquad (11a)$$

with $p_a$ and $R$ both functions of $t', t$ and $z$. Using eqn.9, eqn.11a is rewritten according to:

$$ds_{L,a}(t',t;z) = \frac{p_a}{R}\sqrt{a^2 + (\epsilon^2 - \epsilon)z^2}\,dz. \qquad (11b)$$

The next step is to link $h$ to $z$, through the contact angle. By denoting with $z^*$ the coordinate of the substrate plane, from the equation of the ellipsoid we obtain $z^*(t',t) = kc(t',t)$ with constant $k = \frac{1}{\sqrt{1+(\tan\varphi)^2(1-\epsilon)}}$. It follows that in eqn. 10 $z = h + kc(t',t)$. Moreover, the aspect ratio of the nucleus (ratio between the maximum height and the radius of the substrate/nucleus circular interface) is equal to $Ar = \frac{c-z^*}{(a^2+(\epsilon-1)z^{*2})^{1/2}} = \sqrt{\frac{1-k}{(1+k)(1-\epsilon)}}$ that is constant with time. The differential of eqn.11b is taken at fixed $t$ and $t'$, which entails $dz = dh$. Because of $\partial_t s_a = p_a \partial_t R$ eqn.8 provides, $p_a = P_c \frac{\partial_t s}{\partial_t R} = 2\pi R P_c$ and eqn.11b becomes

$$ds_{L,a}(t',t;h) = 2\pi P_c(t|t';h)\sqrt{a^2(t',t)+(\epsilon^2-\epsilon)[h+kc(t',t)]^2}\,dh \;. \qquad (12a)$$

Using $a^2(t',t) = (1-\epsilon)c^2(t',t)$, the lateral surface of the deposit is eventually given by

$$S_L(t) = 2\pi \int_0^{h_{max}} dh \int_0^{\bar{t}(t,h)} I_a(t')\sqrt{(1-\epsilon)c^2(t',t)-\epsilon(1-\epsilon)[h+kc(t',t)]^2}\,P_c(t|t';h)dt' \qquad (12b)$$

where $h_{max} = c(t,0)(1-k)$ and $\bar{t}(t,h)$ is defined by the equation $c(\bar{t},t) = \frac{h}{1-k}$.

*2.3 Application to KJMA compliant transformations*

*2.3.1 Homogeneous Nucleation*

In the case of transformations consistent with the Kolmogorov-Johnson-Mehl-Avrami (KJMA) model, the distribution of nuclei is random throughout the space, $I_a(t) = I(t)\big(1 - \xi(t)\big)$ and $q = 1$ since the overgrowth phenomenon is precluded. Eqns.3, 5a give

$$P_c(t|t') = \frac{1 - \xi(t)}{1 - \xi(t')},\tag{13}$$

that is the probability that a generic point, which is untransformed at time $t' < t$, is also untransformed at time $t$. This probability is independent of the nucleation event which takes place at $(r(t', t), t')$. Under the same nucleation conditions and parabolic growth, the overgrowth process reduces this probability by a factor of $q < 1$ [15]. However, as demonstrated in ref.[17] the overgrowth phenomenon has a negligible effect on the kinetics and the KJMA model can be safely employed in this case as well. Moreover, parabolic growth usually implies non-random distribution of nuclei owing to the diffusion layer, around growing nuclei, where nucleation is inhibited. It has been shown that under typical conditions of transformations this effect is negligible [18].

The extension of the interface (per unit *D*-volume) becomes

$$X_L(t) = [1 - \xi(t)]X_{L,ex}(t)\,,\tag{14}$$

where the measure of the extended interface is $X_{L,ex}(t) = \int_0^t I(t')x(t', t)dt'$. In the KJMA model

$$\xi(t) = 1 - e^{-V_{ex}(t)}\tag{15a}$$

with $V_{ex}(t) = \int_0^t I(t')v(t',t)dt'$. It follows that $\frac{dV_{ex}(t)}{dt} = \int_0^t I(t')\partial_t v(t',t)dt' = \int_0^t I(t')x(t',t)\partial_t r(t',t)dt'$. For linear growth, $r(t) = \beta t$, we get $\frac{dV_{ex}(t)}{dt} = \beta X_{L,ex}(t)$ and eqn.14 gives $X_L(t) = \frac{1}{\beta}e^{-V_{ex}(t)}\frac{dV_{ex}(t)}{dt} = \frac{1}{\beta}\frac{d\xi(t)}{dt}$.

In the general case of spherical nuclei with power growth, $r(t) = (\beta t)^n$, and constant nucleation rate, eqn.14 provides

$$X_L(t) = g_D e^{-V_{ex}(t)}[V_{ex}(t)]^{\frac{n(D-1)+1}{nD+1}} , \qquad (15b)$$

with $g_D = \left(\frac{I\Omega_D}{\beta}\right)^{\frac{n}{nD+1}}\frac{[D(nD+1)]^{\frac{n(D-1)+1}{nD+1}}}{n(D-1)+1}$ and $V_{ex}(t) = \frac{I\Omega_D}{[D(nD+1)]}\beta^{nD}t^{nD+1}$. For $n = 1$ the result above reported is attained. Setting $\frac{d\xi(t)}{dt} = X_L w$, where $w$ has the meaning of mean growth rate [19], eqns. 15a-b provide the following scaling for $w$

$$w(\xi) \sim [-\ln(1-\xi)]^{\frac{n-1}{nD+1}}. \qquad (15c)$$

For parabolic growth ($n = 1/2$) the power exponents are $-0.25$ and $-0.2$ for 2D and 3D transformations. As discussed below, even in this case the $w$ term can be taken constant, approximately.

### 2.3.2 Heterogeneous Nucleation

In case of heterogeneous nucleation we employ eqn.13 on each layer according to

$$P_c(t|t';h) = \frac{1-S(t,h)}{1-S(t',h)} = e^{-[\sigma_{ex}(t,h)-\sigma_{ex}(t',h)]} , \qquad (16a)$$

where $S(t,h) = 1 - e^{-\sigma_{ex}(t,h)}$ with

$$\sigma_{ex}(t,h) = \pi(1-\epsilon)\int_0^{\bar{t}(t,h)} I(t')[c^2(t',t) - [h + kc(t',t)]^2]dt' \ . \qquad (16b)$$

Since $I_a(t) = I(t)\big(1 - S(t,h)\big) = I(t)e^{-\sigma_{ex}(t,h)}$, for power growth, $c(t) = (\beta t)^n$, eqns.12,16 provide

$$\sigma_{ex}(t,\eta) = \pi(1-\epsilon)\beta^{2n}t^{2n+1}\int_0^{1-\eta} I(\tau')[(1-\tau')^{2n} - [(1-k)\eta^n + k(1-\tau')^n]^2]d\tau' \qquad (17a)$$

and

$$S_L(t) = 2\pi n(1-k)\sqrt{1-\epsilon}\,\beta^{2n}t^{2n+1} \times$$

$$\int_0^1 \eta^{n-1}e^{-\sigma_{ex}(t,\eta)}\,d\eta \int_0^{1-\eta} I(\tau')\sqrt{(1-\tau')^{2n} - \epsilon[(1-k)\eta^n + k(1-\tau')^n]^2}d\tau' \ , \qquad (17b)$$

where the change of variables $\tau' = \frac{t'}{t}$ and $\eta = \frac{1}{\beta t}\left(\frac{h}{1-k}\right)^{1/n}$ has been done.

In the following, we evaluate eqn.17b for some specific cases useful in film growth.

- *Simultaneous nucleation*, $I(t') = N_0\delta(t')$, where $N_0$ is the nucleation density. Eqns.17a-b, provide

$$\sigma_{ex}(t,\eta) = \pi(1-\epsilon)N_0\beta^{2n}t^{2n}[1 - [(1-k)\eta^n + k]^2] \qquad (18a)$$

$$S_L(t) = 2\pi n(1-k)\sqrt{1-\epsilon}\,N_0\beta^{2n}t^{2n} \times$$

$$\int_0^1 \eta^{n-1}e^{-\sigma_{ex}(t,\eta)}\sqrt{1 - \epsilon[(1-k)\eta^n + k]^2}d\eta \qquad (18b)$$

which reduces to the solution of ref.[7] for hemispherical nuclei and linear growth:

$$S_L(t) = 2\sigma_{ex}(t,0) \int_0^1 e^{-\sigma_{ex}(t,0)[1-\eta^2]} d\eta \qquad (18c)$$

with $\sigma_{ex}(t,0) = \pi N_0 \beta^2 t^2$.

- *Spherical-cap nuclei and constant nucleation rate*, $\epsilon = 0$, $k = \cos\varphi$. Eqn.17b gives

$$\sigma_{ex}(t,\eta) = \sigma_{ex}(t,0) \left[ 1 + \eta^{2n+1} \left( \frac{2n(n+1) - 2n^2\cos\varphi}{(1+\cos\varphi)(n+1)} \right) - \frac{\cos\varphi}{(1+\cos\varphi)} \frac{2(2n+1)}{n+1} \eta^n \right.$$
$$\left. - \frac{(2n+1)(1-\cos\varphi)}{(1+\cos\varphi)} \eta^{2n} \right], \qquad (19a)$$

$$S_L(t) = \frac{2n(2n+1)}{(n+1)(1+\cos\varphi)} \sigma_{ex}(t,0) \int_0^1 e^{-\sigma_{ex}(t,\eta)} (\eta^{n-1} - \eta^{2n}) d\eta. \qquad (19b)$$

where $\sigma_{ex}(t,0) = \pi \frac{It^{2n+1}\beta^{2n}\sin^2\varphi}{2n+1} = -\ln(1 - S(t,0))$ with $S(t,0)$ fraction of substrate surface covered by the film. It follows that the interface extension can be expressed as a function of the fraction of substrate surface covered by islands.

For hemispherical nuclei and constant nucleation rate ($\varphi = 1/2$) eqn.19b gives

$$S_L(t) = \frac{2n(2n+1)}{n+1} \sigma_{ex}(t,0) \int_0^1 e^{-\sigma_{ex}(t,0)[1+2n\eta^{2n+1}-(2n+1)\eta^{2n}]} (\eta^{n-1} - \eta^{2n}) d\eta. \qquad (19c)$$

For $n = 1$ eqn.19c reduces to the solution already obtained in ref.[7].

Of interest is also the diffusion-controlled growth of hemispherical nuclei, which implies parabolic growth. Eqn.17b gives ($n = \frac{1}{2}$, $\varphi = \frac{\pi}{2}$)

$$S_L(t) = \frac{4}{3} \sigma_{ex}(t,0) \int_0^1 e^{-\sigma_{ex}(t,0)(1-\eta)^2} (1 - \eta^{3/2}) \frac{1}{\sqrt{\eta}} d\eta. \qquad (20)$$

For the sake of completeness and in view of its relevance for electrochemistry, in the Appendix eqn.17b has been evaluated for exponential nucleation rate [20] and either parabolic or linear growth of hemispherical nuclei. It is shown that the case of exponential nucleation embodies transformations ruled by both simultaneous and progressive nucleation, as limiting cases.

### 2.4 Numerical results

The behavior of the interface evolution for homogeneous nucleation in 2D and 3D space are displayed in Figs.2a,b as a function of the extension of the new phase ($S$ and $V$), for several values of the growth exponent. The functions have the typical bell shape with the maximum depending on growth exponent. In the figure, the interface evolution for simultaneous nucleation is also reported as dashed line, which is independent of growth exponent when expressed in terms of either $S$ or $V$. In fact, in this case eqn.14 provides

$$X_L(X) = G_D(1-X)\left[\ln\frac{1}{1-X}\right]^{\frac{(D-1)}{D}} \tag{21}$$

with $G_D = (\Omega_D N_0 D^{(D-1)})^{1/D}$. For $D=2$ and $D=3$ eqn.21 has been previously obtained in refs.[21, 22, 23] and its validity in the discrete case with von Neumann neighborhoods investigated in refs.[24,25]. The maximum of $X_L$ is attained for $X^* = 1 - e^{-(D-1)/D}$, that is equal to 0.39 and 0.49 for 2D and 3D transitions, respectively. The mean growth rate $w$, for constant nucleation rate is displayed in Fig.2c as a function of the transformed surface (volume) in the case of parabolic growth ($n=1/2$). Apart from the initial stage of the transformation, the variation of $w$ is quite modest and can be considered approximately constant in the central portion of the kinetics.

The evolution of the lateral surface of the interface for heterogeneous nucleation has been computed for progressive and simultaneous nucleation as a function of growth exponent and contact angle of spherical caps. Typical behavior of the surface is displayed in Fig.3a as a function of $\sigma_{ex}(t,0)$, for progressive nucleation at $n=1/2$ and for several values of the contact angle, $\varphi$. In the inset of Fig.3a the behavior of $S_L$ and $S$ at maximum are shown as a function of contact angle for linear and parabolic growths. It stems that $S_L$ and $S$ at maximum are mainly dependent on contact angle, being their variation negligible when shifting from linear to parabolic growth.

The behavior of the interface extension can be highlighted by using normalized form of the kinetics, namely plotting $S_L/S_{L,mx}$ vs $t/t_{mx}$, where $mx$ denotes values at maximum. This is shown in Fig.3b for the curves of Fig.3a. As the contact angle decreases, the maximum becomes less sharp, the kinetics resemble the trend of the extension of the deposit/substrate interface. The comparison between the $n = 1$ and $n = 1/2$ growth regimes is reported in Fig.3c for two values of the contact angle. A decrease of the growth exponent implies lower (greater) values of $t/t_{mx}$ for $\sigma_{ex}(t, 0)/\sigma_{ex}(t_{mx}, 0)$ lower (greater) than one. This is because $\sigma_{ex}(t, 0) \sim t^{2n+1}$ and $\sigma_{ex}(t_{mx}, 0)$ is nearly independent of $n$ (Fig.3a). Accordingly, the portion of the curve before the maximum corresponds to $t/t_{mx}$ values such that $(t/t_{mx})_{n=1/2} < (t/t_{mx})_{n=1} < 1$, i.e. the curve of Fig.3a shifts towards the origin in this domain. The opposite holds true for the curve in the domain above the maximum. The normalized kinetics for progressive and simultaneous nucleation are reported in Fig.3c for linear and parabolic growths. Incidentally, the curve for the simultaneous nucleation at $n = 1/2$ is close to that for progressive nucleation with linear growth. Numerical results of the interface evolution for exponential nucleation rate do show that the kinetics converge to those of simultaneous and progressive nucleation at $k_i \ll 1$ and $k_i \gg 1$, respectively (see also section 2.5). Finally, from eqn.18 it stems that in the case of simultaneous nucleation of hemispheres the lateral surfaces for $n = 1$ and $n = 1/2$ coincide when expressed in terms of $\sigma_{ex}(t, 0)$ variable.

## 2.5 *Application to experimental data*

In the case of homogeneous nucleation eqn.14 has been successfully employed for describing the interface evolution in 2D space. In particular, it has been applied to the simultaneous nucleation of diamond via CVD (chemical vapor deposition) on Silicon substrate, and to describe the extension of the TiO$_2$ HTM interface [8,26]. As far as the 3D case is concerned, for interface controlled growth eqn.15b provides $\frac{dV}{dt} \propto X_L(t) = g_3(1 - V)[-\ln(1 - V)]^{3/4}$ that can be employed to model differential scanning calorimetry (DSC) data [27, 28]. In addition, the evolution of the interface can be useful to characterize the microstructure of alloys during recrystallization. This topic has been discussed in a certain detail in refs.[19, 29] where the concept of "path of microstructural evolution" has been introduced. Among the indicators defined in [19], we consider the "partial path function", $S_L$, which is assumed to be a function of transformed volume, only: $S_L = S_L(V)$. The modeling presented above conforms to this requirement and can be employed to interpret experimental data. An example is displayed in Fig.4a,b for the behavior of $S_L(V)$ during recrystallization of Aluminium alloy and Brass at several temperatures (from refs. [19], [30]). Data points have been normalized to the same value for all temperatures and the behavior is approximately independent of the thermal history of the sample. It was also found that in these alloys recrystallization occurs by simultaneous nucleation [19, 30]. Accordingly, in Fig.4 the interface evolution has been modeled through eqn.21 at $D = 3$. It is worth pointing out that for simultaneous nucleation the $S_L = S_L(V)$ function is independent of growth law of nuclei. Therefore, Eqn.21 can equally be employed when the growth exponent changes during the transformation. The system considered in Fig.4a is significant in this context since the growth exponent is not constant during the entire recrystallization process [19]. Nevertheless, the $S_L = S_L(V)$ curve of eqn.21 can be a suitable path function for this transformation as well.

As regards the evolution of the interface in heterogeneous nucleation, it can be detected through experiments on reaction kinetics occurring at the surface of the deposit. The paradigmatic cases we consider here are the electrodeposition of a new phase at a foreign substrate and the oxide formation by ALD (atomic layer deposition).

When charge transfer is rate-determining, electrodeposition of the new phase is proportional to the lateral surface of the deposit. Therefore, if the deposition occurs by nucleation and growth, the time dependence of the electric current, under potentiostatic conditions, is proportional to the kinetics of the surface evolution with $n = 1$ [7] (and references therein). The model discussed in section 2.3 has been employed for describing the electrodeposition of Zinc on glassy carbon electrode of ref.[31]. The kinetics for simultaneous and progressive nucleation at $n=1$ and $\varphi = 30°$, and $\varphi = 90°$, are found in good agreement with the experimental data as shown in Fig.5. Electrochemical nucleation and growth of Co onto glassy carbon and of Co-Ni alloy onto Cu layer have been investigated through potentiostatic experiments in refs. [5,32]. These systems share similarities as in both of them electrodeposition is kinetic-controlled and the nucleation process is simultaneous. On this basis, normalized experimental data have been modeled by means of eqn.18 for cap-shaped nuclei, and compared to experimental data in Fig.6.

Oxide formation by atomic layer deposition has attracting attention in recent years as a promising technique for growing thin oxide layer on III-V materials. This process open up the possibility to insert III-V compounds into the metal-oxide-semiconductor channel of FET devices [33]. A comprehensive study of ZnO growth at InGaAs substate by ALD has been presented in ref.[10] by using diethylzinc $Zn(C_2H_5)_2$ as the Zn precursor and $H_2O$ as oxidant. After an induction time in which desorption of Zn occurs from the surface, the reaction takes place at the surface of the oxide; the growth kinetics is therefore related to the extension of the lateral surface of the oxide. In ref.[10] the oxide growth has been measured through X-ray fluorescence spectroscopy. The kinetics exhibits a maximum and reach a plateau at longer times. This behavior is ascribed to the roughness of the surface that reaches a maximum before coalescence among nuclei becomes significant. On this basis, a kinetic model has been developed to describe the data, considering an ordered distribution of nuclei, equal in size, on a square lattice [10].

We employed the present approach for describing the ALD data of ref.[10] for three different temperatures. In this work, experimental data on oxide thickening were recorded as a function of the number of cycles that is proportional to time. Once plotted in normalized form, all the kinetics collapse on the same curve as Fig.7 does show, where the induction period was not considered in data analysis. A satisfactory agreement between the model and the experimental data has been attained for spherical nuclei and simultaneous nucleation. It is worth stressing that this nucleation mode is in agreement with that considered in ref.[10] to fit the data.

### 3-Conclusions

We developed a theoretical approach for the kinetics of interface extension in transformations ruled by nucleation and growth. The approach applies to the general case of correlated nucleation and requires the knowledge of the $P_c(t|t')$ conditional probability in terms of correlation functions. The random case was discussed in detail for both homogeneous and heterogeneous nucleation. In the former case an analytical solution is determined which is manageable for describing experimental data. The latter case is more involved providing numerical solution of the kinetics as a function of contact angle and growth exponent. The approach is shown to be suitable for describing experimental data on recrystallization, electrodeposition and ALD by using normalized representation of the kinetics. However, attention must be paid since simultaneous and progressive nucleation may lead to similar normalized curve depending on growth law.

**Appendix A**

To tackle the process of transformation with correlated nuclei, we made use of the theory of stochastic processes in terms of correlation functions. The approach has been discussed in previous works; here we briefly summarize the main results useful for the development of the present modeling. For transformations taking place by homogeneous nucleation and growth the probability that a point of the space belongs to the untransformed phase is given in terms of actual nucleation rate by the expressions in terms of either $f$-functions or correlation functions [13]

$$1 - \xi(t) = 1 + \sum_{k=1}^{\infty} \frac{(-)^k}{k!} \int_0^t dt_1 \, I_a(t_1) \int_0^t dt_2 \, I_a(t_2) \dots$$
$$\times \int_0^t dt_k \, I_a(t_k) \int_{\Delta_{1t}} d\boldsymbol{r}_1 \int_{\Delta_{2t}} d\boldsymbol{r}_2 \dots \int_{\Delta_{kt}} d\boldsymbol{r}_k f_k(\boldsymbol{r}_1, \boldsymbol{r}_2, \dots, \boldsymbol{r}_k) \qquad (A1)$$

$$1 - \xi(t) = \exp\left[ \sum_{k=1}^{\infty} \frac{(-)^k}{k!} \int_0^t dt_1 \, I_a(t_1) \int_0^t dt_2 \, I_a(t_2) \dots \right.$$
$$\left. \times \int_0^t dt_k \, I_a(t_k) \int_{\Delta_{1t}} d\boldsymbol{r}_1 \int_{\Delta_{2t}} d\boldsymbol{r}_2 \dots \int_{\Delta_{kt}} d\boldsymbol{r}_k g_k(\boldsymbol{r}_1, \boldsymbol{r}_2, \dots, \boldsymbol{r}_k) \right], \qquad (A2)$$

where $\xi(t)$ is the ratio of the transformed volume to the whole volume where the transition occurs. In eqn.A1, $[\, I_a(t_1)dt_1 d\boldsymbol{r}_1 ][\, I_a(t_2)dt_2 d\boldsymbol{r}_2 ] \dots [\, I_a(t_k)dt_k d\boldsymbol{r}_k ] f_k(\boldsymbol{r}_1, \boldsymbol{r}_2, \dots, \boldsymbol{r}_k)$ is the probability of finding nuclei born between $t_i$ and $t_i + dt_i$ in the volume elements $d\boldsymbol{r}_i$ around $\boldsymbol{r}_i$, irrespective of the location of other $N - m$ nuclei. In the same equation, the measure of the integration domain, $\Delta_{it}$, is equal to the extended volume of nucleus $v(t_i, t)$. By using the cluster expansion of the $f$-functions in terms of correlation-function, eqn.A1 leads to eqn.A2 where $g_k(\boldsymbol{r}_1, \boldsymbol{r}_2, \dots, \boldsymbol{r}_k)$ is the $k$-nuclei correlation function for nuclei located at the stated positions. Also, since the system is assumed to be translationally invariant, $f_1 = g_1 = 1$ and $f_k = f_k(\boldsymbol{r}_2', \dots \boldsymbol{r}_k')$, $g_k = g_k(\boldsymbol{r}_2', \dots \boldsymbol{r}_k')$ with $\boldsymbol{r}_i' = \boldsymbol{r}_i - \boldsymbol{r}_1$ relative coordinates. Since eqns.A1, A2 are expressed in terms of actual nucleation rate and being actual nuclei spatially correlated, it follows that $f_{k>1} \neq 1$ and $g_k \neq 0$. Noteworthy, this also holds in the case of KJMA compliant transformations where position of actual nuclei is not random; in fact an actual nucleus is constrained to form in the untransformed region. To restore randomness nucleation has to be allowed in the entire space, including the transformed phase, with the phantom-included rate $I = \frac{I_a}{1 - \xi(t)}$. Consequently, in eqns.A1, A2 $f_k = 1$ and $g_{k>1} = 0$ provided $I$ is used instead of $I_a$. In this case eqns.A1, A2 reduce to the celebrated

KJMA equation where $\int_0^t dt_k I(t_k) \int_{\Delta_{kt}} d\boldsymbol{r}_k = \int_0^t dt' I(t') v(t',t)$. Inclusion of phantom nuclei implies constraints on the growth law as discussed in details in refs.[14,15].

For correlated nucleation the conditional probability $P_c(t|t')$ is computed from eqn.A1 and eqns.1a,1b. In particular, eqns.1a,1b give the derivative of $\xi$ according to:

$$\frac{d\xi}{dt} = \int_0^t P_c(t|t') I_a(t') r^{D-1}(t',t) \partial_t r(t',t) d\Omega_D dt'. \qquad (A3)$$

By equating eqn.A3 to the derivative of eqn.A1 it is possible to single out the expression of $P_c(t|t')$. Also, a cluster expansion of the series in terms of correlation function provides the result

$$P_c(t|t') = \exp\left[\sum_{m=1}^{\infty} \frac{(-)^m}{m!} \tilde{I}_m(t,t')\right], \qquad (A4)$$

where

$$\tilde{I}_m(t,t') = \int_0^t d\tau_1 I_a(\tau_1) \dots \int_0^t d\tau_m I_a(\tau_m) \int_{\Delta_{1t}} d\boldsymbol{x}_1 \dots \int_{\Delta_{mt}} d\boldsymbol{x}_m \tilde{g}_m(\boldsymbol{x}_1, \boldsymbol{x}_2, \dots \boldsymbol{x}_m, t', \tau_1, \dots \tau_m). \quad (A5)$$

Eqns.A4, A5 were derived using relative coordinates with respect to nucleus located at $\boldsymbol{r}_1$ where $\boldsymbol{x}_i = (\boldsymbol{r}_{i+1} - R(t,t')\hat{\boldsymbol{r}}_1)$. On this basis $\tilde{g}_{m-1}$ is related through cluster expansion to the $g_k$ correlation functions up to $g_m$. Since $\tilde{g}_1(\boldsymbol{x}_1, t', \tau_1) = 1 + g_2(\boldsymbol{x}_1, t', \tau_1)$, the first term in the argument of eqn.A5 reads [13]

$$\tilde{I}_1(t,t') = \int_0^t d\tau_1 I_a(\tau_1) \int_{\Delta_{1t}} d\boldsymbol{x}_1 \tilde{g}_1(\boldsymbol{x}_1, t', \tau_1) = \int_0^t d\tau_1 I_a(\tau_1) \int_{\Delta_{2t}} d\boldsymbol{x}_1 f_2(\boldsymbol{x}_1, t', \tau_1)$$

$$= \int_0^t dt_2 I_a(t_2) \int_{\Delta_{2t}} d\boldsymbol{r}_2 f_2(\boldsymbol{r}_1, \boldsymbol{r}_2, t', t_2)|_{\boldsymbol{r}_1 = R(t,t')\hat{\boldsymbol{r}}_1} , \qquad (A6)$$

where $f_2$ is the two nuclei $f$-function.

## Appendix B

The nucleation rate is given by $I(t) = I_0 e^{-\gamma t}$ [20] and the extended surface eqn.16b, $\sigma_{ex}(t,h) = \pi I_0 \int_0^{\bar{t}} e^{-\gamma t'} [(\beta(t-t'))^{2n} - h^2 ]$, becomes

$$\sigma_{ex}(\tau,\eta)_{n=1/2} = k_0 [\tau(1-\eta) - (1 - e^{-\tau(1-\eta)})] \qquad (A7)$$

for $n = 1/2$ and

$$\sigma_{ex}(\tau,\eta)_{n=1} = k_1 [\tau^2(1-\eta^2) + 2(1 - e^{-\tau(1-\eta)}) - 2\tau(1 - \eta e^{-\tau(1-\eta)})], \qquad (A8)$$

for $n = 1$, where $k_0 = \pi I_0 \frac{\beta}{\gamma^2}$, $k_1 = \pi I_0 \frac{\beta^2}{\gamma^3}$, $\eta = \frac{\gamma h^{1/n}}{\beta \tau}$ and $\tau = \gamma t$ is the reduced time. The interface surface is given through eqns.17a,b as follows

$$S_L(\tau)_{n=1/2} = 2k_0 e^{-\tau} \tau^2 \int_0^1 e^{-\sigma_{ex}(\tau,\eta)} \frac{d\eta}{\sqrt{\eta}} \int_\eta^1 e^{\tau x} \sqrt{x} dx \qquad (A9)$$

and

$$S_L(\tau)_{n=1} = 2k_1 \int_0^1 e^{-\sigma_{ex}(\tau,\eta)} [\tau^2(1 - \eta e^{-\tau(1-\eta)}) - \tau(1 - \eta e^{-\tau(1-\eta)})] d\eta \qquad (A10)$$

with $\sigma_{ex}$ given by eqns.A7,A8. Eqn.A10 can be rewritten in the alternative form

$$S_L(\tau) = 2k_1 \int_0^1 e^{-\sigma_{ex}(\tau,\eta)} \left[ \tau^2(1-\eta) - \tau\left(1 - e^{-\tau(1-\eta)}\right) \right] d\eta$$
$$+ \left[ e^{-\sigma_{ex}(\tau,1)} - e^{-\sigma_{ex}(\tau,0)} \right] \tag{A11}$$

that is

$$S_L(\tau) = 2k_1 \int_0^1 e^{-\sigma_{ex}(\tau,\eta)} \left[ \tau^2(1-\eta) - \tau\left(1 - e^{-\tau(1-\eta)}\right) \right] d\eta$$
$$+ \left[ 1 - e^{-k_1\left[\tau^2 + 2(1-e^{-\tau}) - 2\tau\right]} \right]. \tag{A12}$$

The formulation above embodies transformations ruled by both simultaneous and progressive nucleation. For $k_i \gg 1$ the constant nucleation rate is recovered while for $k_i \ll 1$ that of simultaneous nucleation with $N_0 = I_0/\gamma$.

**Figures**

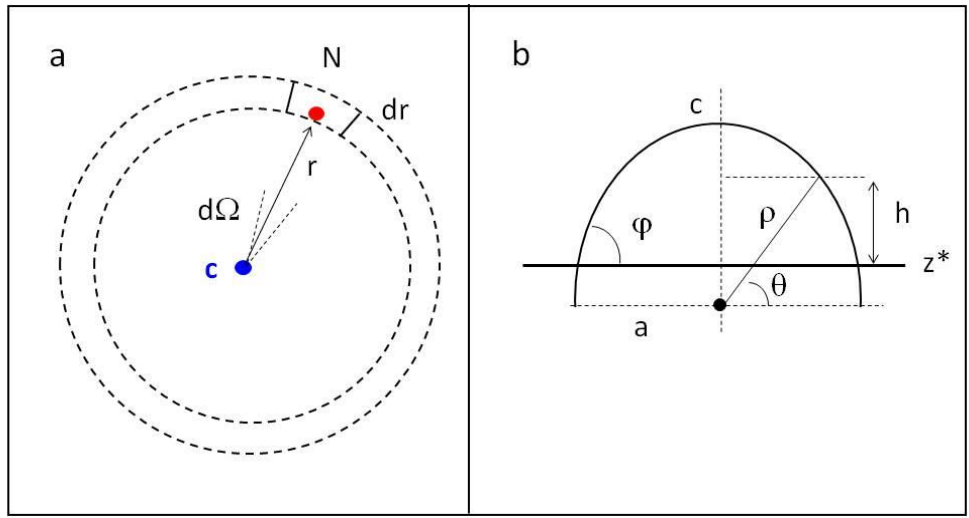

Fig.1

Fig.1 a): Pictorial view of the quantities defined in eqn.1. c is the generic point of the space that is transformed, between $t$ and $t + dt$, by the actual nucleus (N) born between $t'$ and $t' + dt'$ and located at $r = r(t', t)$ within the polar (solid) angle $d\Omega$. b) Definition of the quantities employed in section 2.2.2 for spheroidal nuclei, with semi axes a and c.

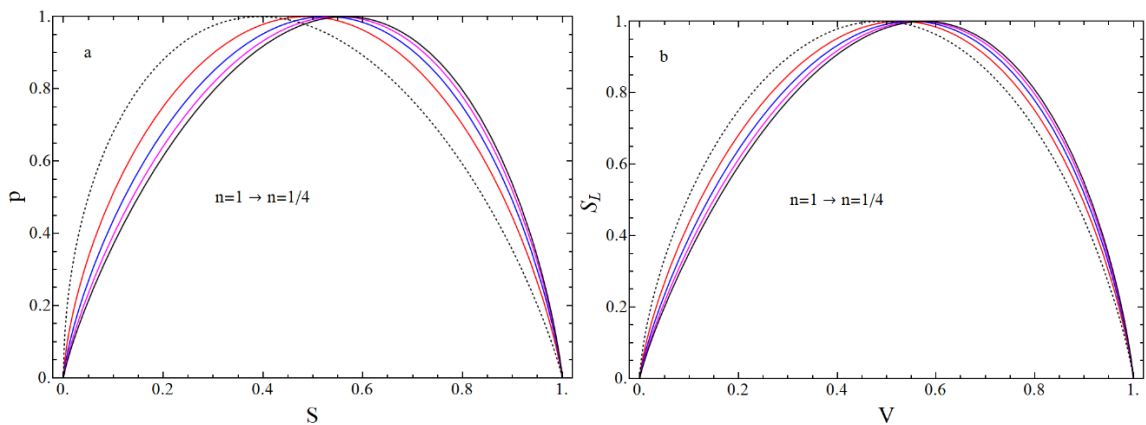

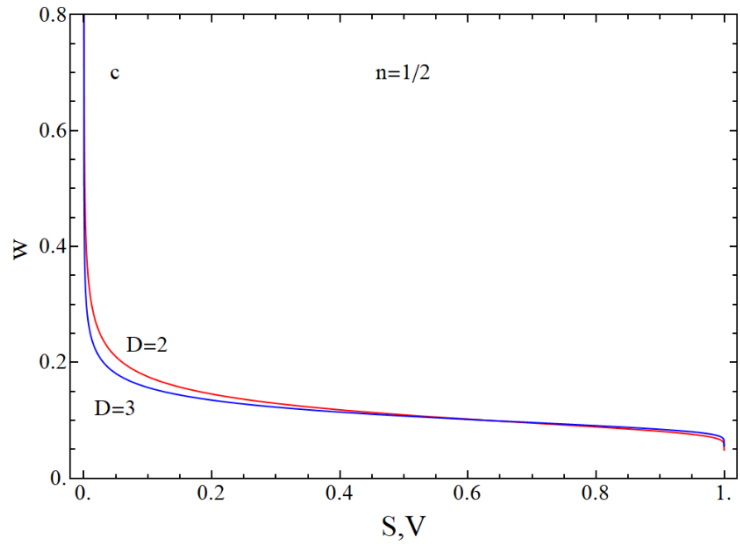

Fig.2

Fig.2 Homogeneous nucleation: Behavior of the interface extension as a function of transformed surface (2D case, panel a) and transformed volume (3D case, panel b). Computations refer to constant nucleation rate and growth exponents $n = 1$, $n = 1/2$, $n = 1/3$ and $n = 1/4$, according to the direction of the arrows. In both panels, dashed lines are the kinetics for simultaneous nucleation. Panel c) displays the mean growth rate, $w$, at $n = 1/2$.

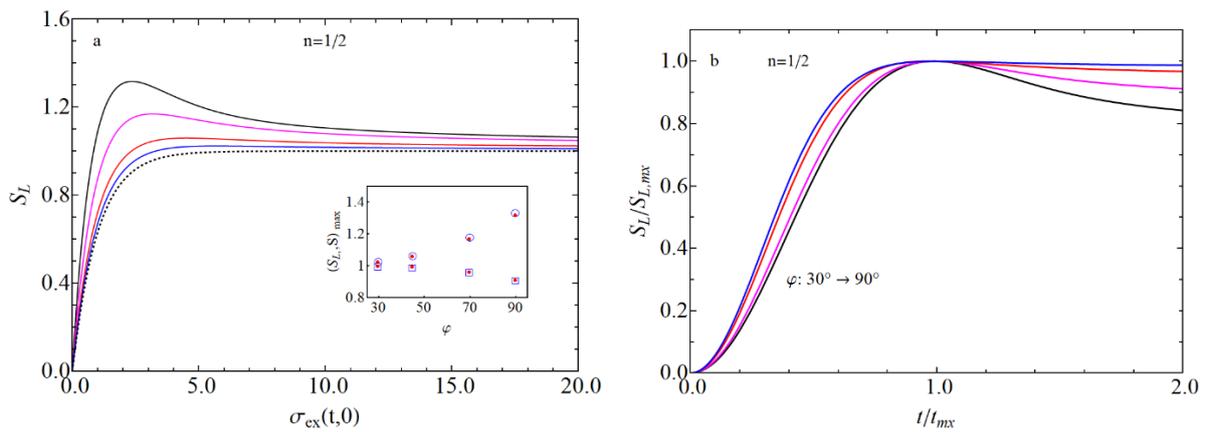

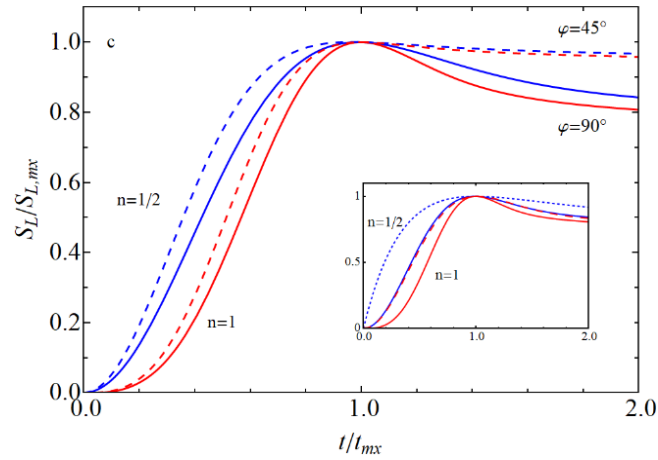

Fig.3

Fig.3 Heterogeneous nucleation.

Panel a). Behavior of the lateral surface of the film as a function of extended surface, for several values of contact angle. Numerical output is for constant nucleation rate and parabolic growth. From the black curve on top: $\varphi = 90°, 70°, 45°, 30°$. The dashed line is the fraction of substrate surface covered by the deposit. Inset: behavior of the coordinates of the maximum of lateral surface, $(S_L, S)_{mx}$, with contact angle. Upper and lower curves are the $S_{L,mx}$ and $S_{mx}$ functions, respectively. Open symbols and full symbols refer to $n = 1$ and $n = 1/2$, respectively.

Panel b). Representation of the curves of panel a) in terms of dimensionless quantities, namely $\frac{S_L}{S_{L,mx}}$ and $\frac{t}{t_{mx}}$.

Panel c) Comparison between linear and parabolic growth in the case of constant nucleation rate. The blue colored curves refer to $n = 1/2$ and the red ones to $n = 1$. Computations for contact angles of $\varphi = 90°$ and $\varphi = 45°$, are displayed as solid and dashed lines, respectively. Inset: Comparison between progressive (solid line) and simultaneous (dashed lines) nucleation at $\varphi = 90°$. The blue colored curves are for $n = 1/2$ and the red ones for $n = 1$.

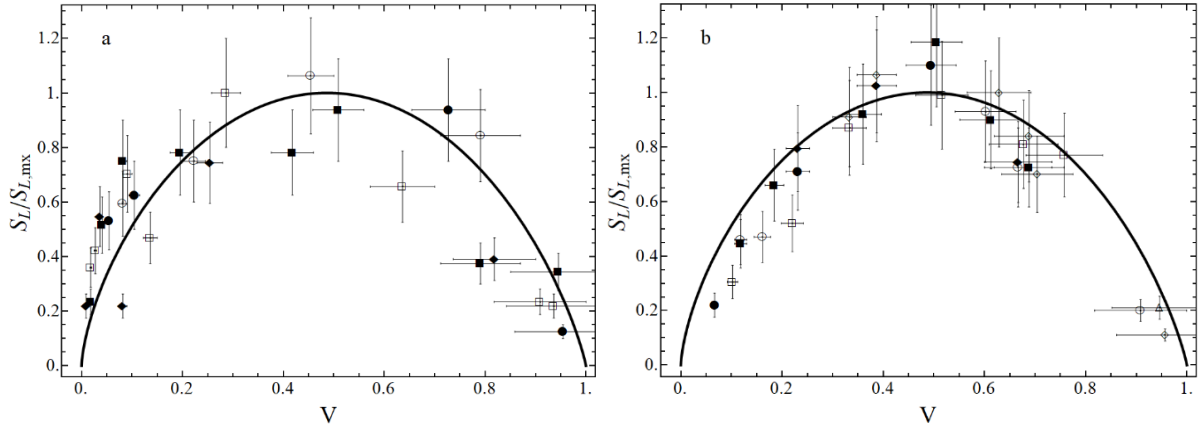

Fig.4

Fig.4 Behavior of the partial path function, $S_L(V)$. Panel a: experimental data on Al-Alloy recrystallization at several temperatures [19]. The normalization factor was the same for all data points. Open squares, $T = 245°C$ (in air); solid diamonds, $T = 245°C$; solid squares, $T = 250°C$; solid circles, $T = 265°C$; open circles, $T = 280°C$. Panel b: experimental data on Brass recrystallization at several temperatures [30]. Solid squares, $T = 250°C$; solid circles, $T = 300°C$; open circles, $T = 350°C$; solid diamonds, $T = 525°C$; open squares, $T = 570°C$, open diamonds, $T = 645°C$ ; open triangles $T = 670°C$. Solid line is the normalized curve given by eqn.21. Error bars show an uncertainty of 10% for both $V$ and $S$.

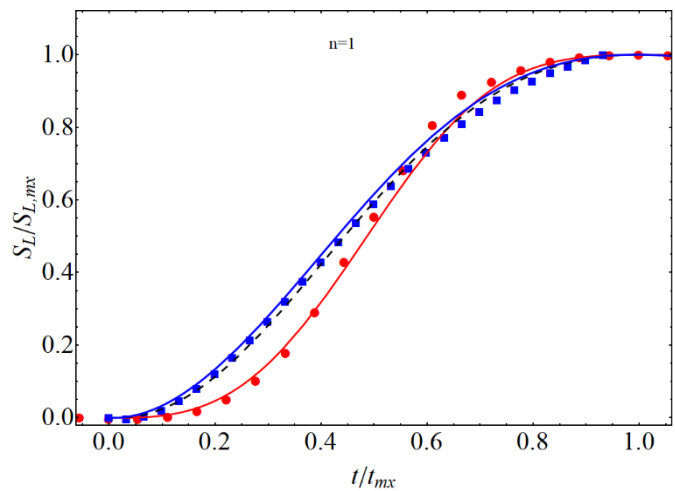

Fig.5

Fig.5 Comparison between theory and experimental data on electrodeposition of Zn on grassy carbon (from ref.[31] ; Squares: potential value -1.07; circles potential value -1.05). Red and Blue curves are the computations for progressive ($\varphi = 30°$ ) and simultaneous ($\varphi = 90°$) nucleation, respectively, at $n = 1$. The dashed line is the kinetics for the exponential nucleation at $\varphi = 90°$ and $k_1 = 0.001$. In the limit of small $k_1$ the exponential nucleation actually reduces to the simultaneous one.

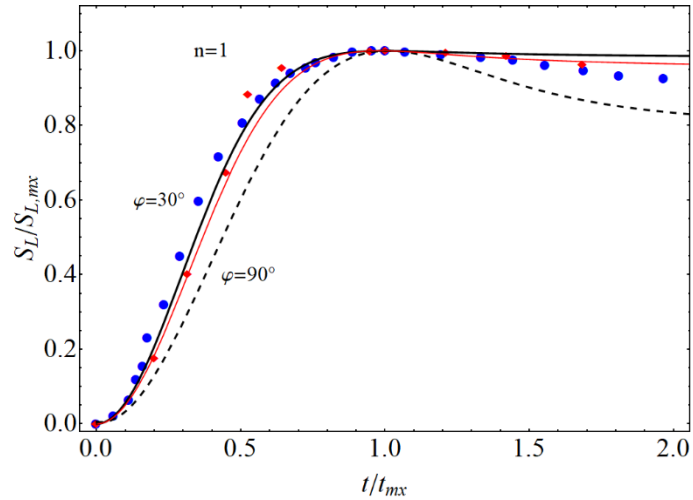

Fig.6

Fig.6 Experimental data on Co (circles) and Co/Ni (diamonds) electrodeposition onto glassy carbon and Cu electrodes [5, 32]. The solid lines are the output of the numerical computations for simultaneous nucleation at $\varphi = 30°$ (black curve), $\varphi = 45°$ (red curve) and for hemispherical nuclei (dashed line).

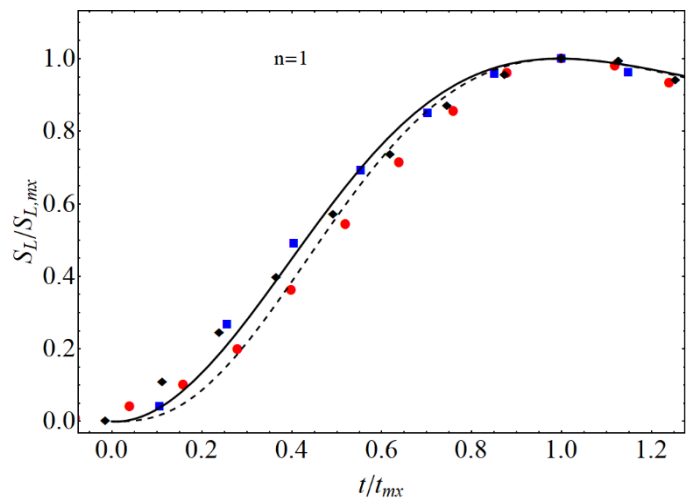

Fig.7

Fig.7 Experimental data on atomic layer deposition from ref.[10] ( $T = 100, 120, 140°C$). When expressed in terms of reduced quantities, the data points at different $T$ lay, nearly, on the same curve.

The computation for simultaneous nucleation of hemispherical nuclei is shown as solid line. Dashed line is the result for the exponential nucleation rate with $k_1 = 0.001$.

# References

1 ) D. Delgadoa , R. Sanchísb, J.A. Ceciliac, E. Rodríguez-Castellónc, A. Caballerod, B. Solsonab, J.M. López Nietoa, Catalysis Today, **333** (2019) 10-16

2 ) J. Zhang, X. Liu , G. Neri , N. Pinna,   Adv. Mater. **28** (2016) 795–831

3 ) N. Yamazoe, Sensors and Actuators B **5** (1991) 7-19

4 ) T. Fujitani, M. Saito, Y. Kanai, T. Kakumoto, T. Watanabe, J. Nakamura, T. Ucijima, Catalysis Letters, **25** (1994) 271-276

5 ) D. Kong, Z. Zheng, F. Meng, N. Li, D. Li, J. of the Electrochem. Soc. **165** (2018) D783-D789

6 )V. Sáez, E. Marchante, M.I. Díez, M.D. Esclapez, P. Bonete, T. Lana-Villarreal, J. González García, J. Mostany, Mater. Chem. And Phys. **125** (2011) 46–54

7 )V.A. Isaev, O.V. Grishenkova, Y.P. Zaykov, J. of Solid State Electrochemistry, **818** (2018) 265

8 ) D. Moia, U. B. Cappel, T. Leijtens, X. Li,  A. M. Telford, H. J. Snaith, B. C. O'Regan,  J. Nelson, P. R. F. Barnes, The Journal of Physical Chemistry C, **119**, (2015)  18975−18985

9 )K.Weinberg, Proc. Appl. Math. Mech. **8**, (2008) 10249-10250

10 )E. V. Skopin, L. Rapenne, H. Roussel, J. L. Deschanvres, E. Blanquet, G. Ciatto, D. D. Fong, M.I. Richard , H. Renevier, Nanoscale **10** (2018) 11585-11596

11 ) A.N. Correia, S.A.S. Machado, J. Braz.Chem. Soc. **8**  (1997)  71-76

12 ) E. Bosco, S. Rangarajan, J. Electroanal. Chem. **134**, (1982) 213

13 ) M. Tomellini, M. Fanfoni, Phys. Rev. E, **90** (5) (2014) 052406

14 ) M. Tomellini, M. Fanfoni, Physical Review B **55** (1997) 14071-3

15 ) N.V. Alekseechkin, J. Non. Cryst. Solids **357** (16-17) (2011) 3159-316

16 ) M. Tomellini, S. Politi, Physica A**513** (2019) 175

17 ) M.P. Shepilov, Glass Physics and Chemistry, **30** (4) (2004) 292-299; M.P. Shepilov, Crystallogr. Rep. **50** (3) (2005) 513-516

18 ) E. Pineda, T. Pradell, D. Crespo, Phil. Magazine A **82** (1) (2002)  107-121

19 ) R. A. Vandermeer, D.J. Jensen, Acta Mater. **49** (11)  (2001) 2083-2094

20 ) M. Palomar-Pardavé, B.R. Scharifker, E.M. Arce, M. Romero-Romo, Electrochem. Acta **50** (24) (2005) 4736-4745

21 ) M. Tomellini, M. Fanfoni, Surf. Sci. Lett. **349** (1996)  L191-L198

22 ) P. R. Rios, J. C. P. T. Oliveira, V.T. Oliveira, J. A. Castro, Materials Research **9** (2)  (2006) 165-170

23 ) P. R. Rios, A.F. Padilha, Mater. Research **6** (4) (2003) 605-613

24) A. Korobov, Phys. Rev. E **79** (2009) 031607

25 ) A. Korobov, Phys. Rev. E 89 (2014) 032405

26 ) R. Polini, M. Tomellini, M. Fanfoni, F. Le Normand, Surf. Sci. **373** (1997) 230-23

27 ) F. Liu, F. Sommer, C. Bos, E.J. Mittemeijer, Inter. Mater. Reviews  **52**  (2007) 193-211

28 J. Farjas, P. Roura, Acta Mater. **54**  (2006)  5573-5579

29 ) R. A. Vandermeer, Acta Mater. **53** (5)  (2005) 1449-1457

30 ) D.A. Mehta, G. Krauss, J. Heat Treating, **2** (1) (981) 83-91

31 ) L.H. Mendoza-Huizar, C.H. Rios-Reyes, M.G. Gomez-Villagas, J. Mex. Chem. Soc. **53/4** (2009) 243

32 ) A. Dolati, S.S. Mahshid, Mater. Chem. And Phys. **108** (2008) 391-396

33 ) M. Paladugu, C. Merckling, R. Loo, O. Richard, H. Bender, J. Dekoster, W. Vandervorst, M. Caymax and M. Heyns, Cryst. Growth Des. **12**, (2012) 4696–4702